\documentclass[pre,amsmath,amssymb,reprint]{revtex4-2}
\usepackage{epsfig}                                                                 
\usepackage{dcolumn}                                                                
\usepackage{bm}                                                                     
\let\boldsymbol\pmb
\usepackage[pdfstartview=FitH,bookmarksopen=true,bookmarksopenlevel=1]{hyperref}    

\begin{document}
\title[Dependence of the surface tension and contact angle on the temperature]%
{Dependence of the surface tension and contact angle on the temperature,\protect\\ as described by the diffuse-interface model}
\author{E. S. Benilov}
 \altaffiliation[]{Department of Mathematics and Statistics, University of Limerick, Limerick, V94 T9PX, Ireland}
 \email{Eugene.Benilov@ul.ie}
 \homepage{https://staff.ul.ie/eugenebenilov/}
\date{\today}

\begin{abstract}
Four results associated with the diffuse-interface model (DIM) for contact
lines are reported in this paper. \underline{First}, a boundary condition is
derived, which states that the fluid near a solid wall must have a certain
density $\rho_{0}$ depending on the solid's properties. Unlike previous
derivations, the one presented here is based on the same physics as the DIM
itself and does not require additional assumptions. \underline{Second},
asymptotic estimates are used to check a conjecture lying at the foundation of
the DIM, as well as all other models of contact lines: that liquid--vapor
interfaces are nearly isothermal. It turns out that, for water, they are not
-- although, for a more viscous fluid, they can be. The non-isothermaility
occurs locally, near the interface, but can still affect the contact-line
dynamics. \underline{Third}, the DIM coupled with a realistic equation of
state for water is used to compute the dependence of the surface tension
$\sigma$ on the temperature $T$, which agrees well with the empiric
$\sigma(T)$. \underline{Fourth}, the same framework is used to compute the
static contact angle of a water--vapor interface. It is shown that, with
increasing temperature, the contact angle becomes either $180^{\circ}$
(perfect hydrophobicity) or $0^{\circ}$ (perfect hydrophilicity), depending
on whether $\rho_{0}$ matches the density of saturated vapor or liquid,
respectively. Such behavior presumably occurs in all fluids, not just water,
and for all sufficiently strong variations of parameters, not just that of the
temperature -- as corroborated by existing observations of drops under
variable electric field.
\end{abstract}
\maketitle

\section{Introduction}

The diffuse-interface model (DIM)
\cite{Gouin87,AndersonMcFaddenWheeler98,PismenPomeau00,Jacqmin00} is based on
an assumption that the van der Waals force in fluids can be described by a
pair-wise potential exerted by the molecules on each other. If the potential's
spatial scale is much shorter than that of the flow, the force term in the
governing equations can be simplified, yielding the so-called Korteweg stress
\cite{Korteweg01}. The resulting model provides a tool for studying flows
involving contact lines, i.e., curves where the gas, liquid, and solid are in
simultaneous contact. To this end, one also needs a boundary condition for
fluid--solid interfaces, of which several versions exist in the literature.
Firstly, Ref. \cite{Seppecher96} suggested a condition prescribing the density
gradient in the direction normal to the solid boundary; secondly, Ref.
\cite{PismenPomeau00} put forward a condition prescribing a linear combination
of the density gradient and the density itself. It was also conjectured in
Ref. \cite{PismenPomeau00} that, if the solid--fluid interaction is
short-ranged by comparison with the fluid--fluid one, the general boundary
condition can be simplified, so that just the density is prescribed. This
simplest boundary condition is usually employed in applications (e.g., Refs.
\cite{DingSpelt07,YueFeng11,KusumaatmajaHemingwayFielding16,FakhariBolster17,BorciaBorciaBestehornVarlamovaHoefnerReif19}
and references therein).

Curiously, there is only one work, Ref. \cite{Caupin05}, where the DIM is
coupled with a realistic equation of state (EoS). The one used in most other
papers is inconsistent with the ideal-gas limit and does not involve
temperature (the latter amounts to spatial isothermality). It allows one,
however, to find analytically the profile of the liquid--vapor interface,
which comes handy when calculating the flow's macroscopic characteristics (the
surface tension and contact angle). Still, the use of a non-realistic EoS
renders the DIM somewhat phenomenological rather than physics-based.

A model close to, but still not quite, realistic was examined in Ref.
\cite{Onuki07}, where the DIM was coupled with the van der Waals EoS.
Interestingly, simulations carried out in this work showed that interfacial
flows can be significantly non-isothermal. This conclusion was confirmed in
Ref. \cite{Benilov20} by fitting the van der Waals EoS to several specific
fluids including water and considering the resulting asymptotic models.

The discrepancies associated with non-realistic equations of state are
resolved in the present work. It concentrates on water -- not only because of
the importance of this fluid, but also because its parameters are well
researched, making it easy to verify theoretical results.

In Sects. \ref{Sect. 2}-\ref{Sect. 3}, the simplest version of the boundary
condition (the one conjectured in Ref. \cite{PismenPomeau00}) will be derived
\emph{without} assuming that the solid--fluid interactions are short-ranged by
comparison with the fluid--fluid ones. It is also shown that, if the DIM is
coupled with a realistic EoS for water, it predicts that interfaces are not
isothermal, which confirms the results of Ref. \cite{Benilov20}. In Sects.
\ref{Sect. 4}-\ref{Sect. 5}, the DIM is used to calculate the dependence of
the surface tension and contact angle on the temperature (both for water).

\section{Formulation\label{Sect. 2}}

\subsection{Basic thermodynamics of non-ideal fluids}

Let $\rho$ be the mass density of a fluid, and $s$ and $e$ be the entropy and
internal energy (both per unit mass), respectively. Then, the fluid's
properties are fully determined by the function $e(\rho,s)$; the temperature
$T$ and pressure $p$, for example, are given by%
\begin{equation}
T=\left(  \frac{\partial e}{\partial s}\right)  _{\rho},\qquad p=\rho
^{2}\left(  \frac{\partial e}{\partial\rho}\right)  _{s}, \label{2.1}%
\end{equation}
where, as usual, the subscripts imply that the corresponding variables are
held constant.

Instead of using $s$ as one of the primary thermodynamic variables, it is more
convenient to use $T$. Rewriting the first equality of (\ref{2.1}) in terms of
$\left(  \rho,T\right)  $, one obtains a restriction linking allowable
$e(\rho,T)$ and $s(\rho,T)$,%
\begin{equation}
\left(  \dfrac{\partial e}{\partial T}\right)  _{\rho}=T\left(  \dfrac
{\partial s}{\partial T}\right)  _{\rho}. \label{2.2}%
\end{equation}
Rewriting the second equality of (\ref{2.1}) and taking into account
(\ref{2.2}), one obtains the EoS,%
\begin{equation}
p=-T\rho^{2}\left(  \dfrac{\partial s}{\partial\rho}\right)  _{T}-a\rho^{2},
\label{2.3}%
\end{equation}
where%
\begin{equation}
a=-\left(  \frac{\partial e}{\partial\rho}\right)  _{T}, \label{2.4}%
\end{equation}
can be interpreted as the first van der Waals parameter -- but, unlike its
classical counterpart, it may depend on $\rho$ and $T$.

Introduce also the specific heat capacity%
\[
c_{V}=\left(  \frac{\partial e}{\partial T}\right)  _{\rho}%
\]
and the Gibbs free energy%
\begin{equation}
G=e-Ts+\frac{p}{\rho}. \label{2.5}%
\end{equation}
Using (\ref{2.2})-(\ref{2.3}), one can show that $G$ is related to the
pressure by%
\begin{equation}
\left(  \dfrac{\partial G}{\partial\rho}\right)  _{T}=\frac{1}{\rho}\left(
\frac{\partial p}{\partial\rho}\right)  _{T}. \label{2.6}%
\end{equation}
The general results in this work will be illustrated using the Enskog--Vlasov
(EV) equation of state, resulting from the hydrodynamic approximation of the
EV kinetic equation
\cite{Desobrino67,Grmela71,BenilovBenilov18,BenilovBenilov19} and implying%
\begin{equation}
e=c_{V}T-a\rho,\qquad s=c_{V}\ln T-R\ln\rho-R\,\Theta(b\rho), \label{2.7}%
\end{equation}
where $c_{V}$ and $a$ are independent of $\rho$ and $T$, $R$ is the specific
gas constant, $b$ is the EV equivalent of the second van der Waals parameter,
and $\Theta(\xi)$ (with $\xi=b\rho$) is a fluid-specific function describing
the non-ideal part of the entropy. Substitution of (\ref{2.7}) into
(\ref{2.3}) yields%
\begin{equation}
p=RT\rho\left[  1+b\rho\,\Theta^{\prime}(b\rho)\right]  -a\rho^{2},
\label{2.8}%
\end{equation}
where $\Theta^{\prime}(\xi)=\mathrm{d}\Theta(\xi)/\mathrm{d}\xi$. Note that,
in applications of the EV theory to real fluids
\cite{BenilovBenilov18,BenilovBenilov19}, the best choice for $b$ turned out
to be the reciprocal of the fluid's triple-point density.

Observe that Eq. (\ref{2.8}) includes the van der Waals EoS as a particular
case with $\Theta(\xi)=-\ln\left(  1-\xi\right)  $.

\subsection{The governing equations and boundary conditions}

Traditionally, the diffuse-interface model is introduced through the free
energy of fluid--fluid and solid--fluid interactions \cite{PismenPomeau00}. It
seems simpler, however, to do so through pair-wise forces exerted by the fluid
molecules on each other, and the forces exerted on the molecules by the
(solid) walls.

Let the former forces be described by an isotropic potential $\Phi(r)$ ($r$ is
the distance between the interacting molecules) and the latter, by a potential
$U(\mathbf{r})$ which decays rapidly when $\mathbf{r}$ moves away from the wall.

Introducing the molecular mass $m$ (so that $\rho/m$ is the number density),
one can express the total collective force in the form%
\begin{multline}
\mathbf{F}(\mathbf{r},t)=-\frac{\rho(\mathbf{r},t)}{m}\\
\times\boldsymbol{\boldsymbol{\nabla}}\left[  \int_{\mathcal{D}}\frac
{\rho(\mathbf{r}_{1},t)}{m}\Phi(\left\vert \mathbf{r-r}_{1}\right\vert
)\,\mathrm{d}^{3}\mathbf{r}_{1}+U(\mathbf{r})\right]  , \label{2.9}%
\end{multline}
where $\mathcal{D}$ is the domain occupied by the fluid (physically, the container).

A compressible Newtonian fluid affected by a force $\mathbf{F}$ is governed by
\cite{FerzigerKaper72}%
\begin{equation}
\frac{\partial\rho}{\partial t}+\boldsymbol{\boldsymbol{\nabla}}\cdot\left(
\rho\mathbf{v}\right)  =0, \label{2.10}%
\end{equation}%
\begin{equation}
\frac{\partial\mathbf{v}}{\partial t}+\left(  \mathbf{v}\cdot
\boldsymbol{\boldsymbol{\nabla}}\right)  \mathbf{v}+\frac{1}{\rho
}\boldsymbol{\boldsymbol{\nabla}}\cdot\left(  \mathbf{I}\,p-\boldsymbol{\Pi
}\right)  =\frac{1}{\rho}\mathbf{F}, \label{2.11}%
\end{equation}%
\begin{multline}
\rho c_{V}\left(  \frac{\partial T}{\partial t}+\mathbf{v}\cdot
\boldsymbol{\boldsymbol{\nabla}}T\right)  +\left[  \mathbf{I}\left(
p+a\rho^{2}\right)  -\boldsymbol{\Pi}\right]  :\boldsymbol{\boldsymbol{\nabla
}}\mathbf{v}\\
-\boldsymbol{\boldsymbol{\nabla}}\cdot\left(  \kappa
\boldsymbol{\boldsymbol{\nabla}}T\right)  =0. \label{2.12}%
\end{multline}
where $\mathbf{I}$ is the identity matrix,%
\begin{equation}
\boldsymbol{\Pi}=\mu_{s}\left[  \boldsymbol{\boldsymbol{\nabla}}%
\mathbf{v}+\left(  \boldsymbol{\boldsymbol{\nabla}}\mathbf{v}\right)
^{T}-\frac{2}{3}\mathbf{I}\left(  \boldsymbol{\boldsymbol{\nabla}}%
\cdot\mathbf{v}\right)  \right]  +\mu_{b}\,\mathbf{I}\left(
\boldsymbol{\boldsymbol{\nabla}}\cdot\mathbf{v}\right)  , \label{2.13}%
\end{equation}
is the viscous stress tensor, $\mu_{s}$ ($\mu_{b}$) is the shear (bulk)
viscosity, and $\kappa$, the thermal conductivity.

Observe that the governing equations (\ref{2.9})-(\ref{2.12}) are invariant
with respect to the simultaneous substitution%
\begin{equation}
\Phi(r)=\Phi_{new}(r)+C\,\delta(\mathbf{r}), \label{2.14}%
\end{equation}%
\begin{equation}
p=p_{new}-\frac{C}{2m^{2}}\rho^{2},\qquad a=a_{new}+\frac{C}{2m^{2}},
\label{2.15}%
\end{equation}
where $\delta(\mathbf{r})$ is the Dirac delta-function and $C$ is an arbitrary
constant. Furthermore, recalling (\ref{2.3})-(\ref{2.4}), one can see that
substitutions (\ref{2.15}) both correspond to%
\[
e=e_{new}-\frac{C}{2m^{2}}\rho.
\]
Choosing in (\ref{2.14}) an appropriate value of $C$, one can make $\Phi
_{new}$ satisfy (the subscript $_{new}$ omitted)%
\begin{equation}
\int\Phi(r)\,\mathrm{d}^{3}\mathbf{r}=0, \label{2.16}%
\end{equation}
where integration is to be carried out over the whole space. In what follows,
the so-called Korteweg parameter will be needed, given by%
\begin{equation}
K=-\frac{1}{m^{2}}\int r^{2}\Phi(r)\,\mathrm{d}^{3}\mathbf{r}. \label{2.17}%
\end{equation}
At $\partial\mathcal{D}$ (the container walls), Eqs. (\ref{2.9})-(\ref{2.13})
should be complemented by the no-flow condition,%
\begin{equation}
\mathbf{v}=\mathbf{0}\qquad\text{at}\qquad\mathbf{r}\in\partial\mathcal{D},
\label{2.18}%
\end{equation}
and a boundary condition for the temperature. The latter does not play a role
in this work, so it is not discussed.

Most importantly, the governing equations do \emph{not} require a boundary
condition for the density$\ $(as the term $\mathbf{v}\cdot
\boldsymbol{\boldsymbol{\nabla}}\rho$ in Eq. (\ref{2.10}) vanishes at
$\mathbf{r}\in\partial\mathcal{D}$ due to (\ref{2.18}), and the other
equations do not involve derivatives of $\rho$).

\subsection{Nondimensionalization}

Let $\bar{r}$ be the spatial scale of the flow and $\bar{v}$, its
characteristic velocity, so the time scale is $\bar{r}/\bar{v}$. The density
will be scaled by its triple-point value (denoted by $b^{-1}$, with a view of
using the EV\ EoS later); the pressure will be scaled by $b^{-2}\bar{a}$
[where $\bar{a}\ $is the characteristic value of $a(\rho,T)$]; and the
temperature, by a characteristic value $\bar{T}$.

The following nondimensional variables will be used:%
\[
\mathbf{r}_{nd}=\frac{\mathbf{r}}{\bar{r}},\qquad t_{nd}=\frac{\bar{v}}%
{\bar{r}}t,
\]%
\[
\rho_{nd}=b\rho,\qquad\mathbf{v}_{nd}=\frac{\mathbf{v}}{\bar{v}},\qquad
p_{nd}=\frac{b^{2}}{\bar{a}}p,\qquad T_{nd}=\frac{T}{\bar{T}}.
\]
It is convenient to also introduce the nondimensional versions of the fluid
parameters. Assume for simplicity that the bulk and shear viscosities are of
the same order ($\sim\bar{\mu}$) and denote the other two scales by
$\bar{\kappa}$ and $\bar{c}_{V}$, so that%
\[
\left(  \mu_{s}\right)  _{nd}=\frac{\mu_{s}}{\bar{\mu}},\qquad\left(  \mu
_{b}\right)  _{nd}=\frac{\mu_{b}}{\bar{\mu}},\qquad\kappa_{nd}=\frac{\kappa
}{\bar{\kappa}},
\]%
\[
\left(  c_{V}\right)  _{nd}=\frac{c_{V}}{\bar{c}_{V}},\qquad a_{nd}=\frac
{a}{\bar{a}}.
\]
Then the nondimensional viscous stress is%
\[
\boldsymbol{\Pi}_{nd}=\frac{\bar{r}}{\bar{\mu}\bar{v}}\boldsymbol{\Pi}.
\]
As shown in Sect. \ref{Sect. 4.3} below, the spatial scale of a \emph{static}
interface is%
\[
\bar{r}=\left(  \frac{K}{\bar{a}}\right)  ^{1/2},
\]
which should also apply to a \emph{moving} one. Such a scaling makes the van
der Waals force comparable, but not necessarily equal, to the pressure
gradient. One should also require that the viscous stress be comparable to the
pressure gradient (as done in the lubrication approximation), which implies%
\[
\bar{v}=\frac{\bar{a}\bar{r}}{\bar{\mu}b^{2}}.
\]
Physically, $\bar{v}$ characterizes a flow due to a disbalance between the van
der Waals force and the pressure gradient (typically, resulting from the
interface being curved) -- whereas the global flow can have a very different
velocity scale.

The DIM is based on an assumption that the spatial scale of $\Phi(r)$ is much
smaller than that of the flow: the latter is $\bar{r}$, so let the former be
$\varepsilon\bar{r}$ with $\varepsilon\ll1$. The scale separation allows one
to approximate the fluid--fluid interaction by the so-called Korteweg stress
-- accordingly, it is convenient to scale $\Phi$ using the Korteweg parameter
(\ref{2.17}):%
\[
\Phi(r)=\frac{Km^{2}}{\varepsilon^{5}\bar{r}^{5}}\Phi_{nd}(\varepsilon
^{-1}r_{nd}),
\]
where the factor of $\varepsilon^{5}$ is inserted to make the nondimensional
version of the Korteweg parameter equal unity,%
\begin{equation}
\int\left(  \varepsilon^{-1}r_{nd}\right)  ^{2}\Phi_{nd}(\varepsilon
^{-1}r_{nd})\,\mathrm{d}^{3}(\varepsilon^{-1}\mathbf{r}_{nd})=1. \label{2.19}%
\end{equation}
The solid--fluid potential will be scaled so that the two terms on the
right-hand side of (\ref{2.9}) are comparable -- which amounts to%
\[
U(\mathbf{r})=\frac{Km}{\varepsilon^{5}\bar{r}^{2}b}U_{nd}(\varepsilon
^{-1}\mathbf{r}_{nd}).
\]
In terms of the nondimensional variables, Eqs. (\ref{2.9})-(\ref{2.13}) have
the form (the subscript $_{nd}$ omitted):%
\begin{equation}
\frac{\partial\rho}{\partial t}+\boldsymbol{\boldsymbol{\nabla}}\cdot\left(
\rho\mathbf{v}\right)  =0, \label{2.20}%
\end{equation}%
\begin{multline}
\fbox{$\alpha$}\left[  \frac{\partial\mathbf{v}}{\partial t}+\left(
\mathbf{v}\cdot\boldsymbol{\boldsymbol{\nabla}}\right)  \mathbf{v}\right]
+\frac{1}{\rho}\boldsymbol{\boldsymbol{\nabla}}\cdot\left(  \mathbf{I}%
\,p-\boldsymbol{\Pi}\right) \\
=-\frac{1}{\varepsilon^{5}}\boldsymbol{\boldsymbol{\nabla}}\left[
\int_{\mathcal{D}}\rho(\mathbf{r}_{1},t)\,\Phi(\varepsilon^{-1}\left\vert
\mathbf{r-r}_{1}\right\vert )\,\mathrm{d}^{3}\mathbf{r}_{1}\right. \\
\mathbf{+}\left.  U(\varepsilon^{-1}\mathbf{r}%
)\vphantom{\int_{\mathcal{D}}}\right]  , \label{2.21}%
\end{multline}%
\begin{multline}
\fbox{$\alpha\gamma$}\rho c_{V}\left(  \frac{\partial T}{\partial
t}+\mathbf{v}\cdot\boldsymbol{\boldsymbol{\nabla}}T\right)  +\fbox{$\beta
$}\left[  \mathbf{I}\left(  p+a\rho^{2}\right)  -\boldsymbol{\Pi}\right]
:\boldsymbol{\boldsymbol{\nabla}}\mathbf{v}\\
-\boldsymbol{\boldsymbol{\nabla}}\cdot\left(  \kappa
\boldsymbol{\boldsymbol{\nabla}}T\right)  =0, \label{2.22}%
\end{multline}
where $\boldsymbol{\Pi}$ is still given by (\ref{2.13}) and%
\begin{equation}
\alpha=\frac{K}{\bar{\mu}^{2}b^{3}},\qquad\beta=\dfrac{\bar{a}K}{\bar{\mu}%
\bar{\kappa}\bar{T}b^{4}},\qquad\gamma=\frac{\bar{c}_{V}\bar{\mu}}{\bar
{\kappa}}. \label{2.23}%
\end{equation}
As follows from the positions of $\alpha$ and $\beta$ in Eqs. (\ref{2.21}%
)-(\ref{2.22}), the former is the Reynolds number and the latter is an
`isothermality parameter' controlling the production of heat by
compressibility and viscosity (if $\beta\ll1$, the flow is close to
isothermal). $\gamma$, in turn, is the Prandtl number.

Finally, one can rewrite (\ref{2.16}) and (\ref{2.19}) in the form%
\begin{equation}
\int\Phi(r_{1})\,\mathrm{d}^{3}\mathbf{r}_{1}=0,\qquad\int r_{1}^{2}\Phi
(r_{1})\,\mathrm{d}^{3}\mathbf{r}_{1}=1, \label{2.24}%
\end{equation}
where $\mathbf{r}_{1}=\varepsilon^{-1}\mathbf{r}_{nd}$.

\section{Asymptotic estimates\label{Sect. 3}}

\subsection{The nondimensional parameters}

In this subsection, the nondimensional parameters $\alpha$, $\beta$, and
$\gamma$ will be estimated for water.

Note that `our' $\bar{a}$ and $b$, are similar to, but not the same as, those
in the van der Waals EoS. The latter are defined by fitting the EoS to the
parameters of the critical point, making the result inaccurate at room temperature.

In this work, $\bar{a}$ and $b$ were determined through the Enskog--Vlasov
EoS, which is much more flexible than its van der Waals counterpart. The
details can be found in Appendix \ref{Appendix A}, together with $\bar{a}$ and
$b$ given by (\ref{A.1})-(\ref{A.2}), respectively. The Korteweg parameter
$K$, in turn, is estimated in Sect. \ref{Sect. 4} and given by (\ref{4.7}).

To estimate $\alpha$, $\beta$, and $\gamma$, one also needs the characteristic
heat capacity $\bar{c}_{V}$, viscosity $\bar{\mu}$, and thermal conductivity
$\bar{\kappa}$. In the context of interfacial dynamics, it is reasonable to
determine these parameters as the average of those for liquid and vapor.
Assuming the temperature of $\bar{T}=25~\mathrm{C}^{\circ}$ and using the data
from Sect. 6.1 of Ref. \cite{HaynesLideBruno17}, one obtains%
\[
\bar{c}_{V}=2.7892~\mathrm{kJ~kg}^{-1}\mathrm{K}^{-1},
\]%
\[
\bar{\mu}=449.87~\mathrm{\mu Pa~s},\qquad\bar{\kappa}=312.45~\mathrm{mW~m}%
^{-1}\mathrm{K}^{-1}.
\]
Substituting these values, into (\ref{2.23}), one obtains%
\[
\alpha\approx0.121,\qquad\beta\approx1.234,\qquad\alpha\gamma\approx0.486.
\]
Interestingly, the estimates of the above parameters based on the (much less
accurate) van der Waals EoS \cite{Benilov20} yield comparable values:
$\alpha\approx0.143$, $\beta\approx0.711$, and $\alpha\gamma\approx0.880$.

It is also worth mentioning that, with increasing $\bar{T}$, the Reynolds
number $\alpha$ grows -- i.e., high-temperature interfacial flows may be close to
inviscid. The isothermality parameter $\beta$, in turn, decreases, but never
becomes small -- not even when the temperature approaches its critical value.
For $\bar{T}=360^{\circ}\mathrm{C}$, for example,%
\[
\alpha\approx13.3,\qquad\beta\approx0.581,\qquad\alpha\gamma\approx6.60.
\]
Thus, interfaces in water are generally \emph{non}-isothermal due to the heat
production by viscosity and compressibility. Even though this effect is local
-- i.e., occurs near the interface -- it can strongly affect the dynamics of
contact lines.

In what follows, only moderate (room) temperatures will be considered,
corresponding to the following asymptotic regime:%
\begin{equation}
\alpha\ll1,\qquad\beta\sim1,\qquad\gamma\sim1.\label{3.1}%
\end{equation}
Other regimes, arising for other fluids, have been examined in Ref.
\cite{Benilov20} using the van der Waals EoS. Eight fluids were considered
(acetone, benzene, ethanol, ethylene glycol, glycerol, mercury, methanol, and
water), and only for ethylene glycol and glycerol $\beta$ has turned out to be
small. Thus, non-isothermality of liquid--vapor interfaces is likely to be a
rule rather than an exception.

\subsection{The asymptotic equations}

The \underline{density} equation (\ref{2.20}), does not involve any parameters
and, thus, remains as is.

Assuming limit (\ref{3.1}) and omitting small terms from the
\underline{temperature} equation (\ref{2.22}), one obtains%
\[
\beta\left[  \mathbf{I}\left(  p+a\rho^{2}\right)  -\boldsymbol{\Pi}\right]
:\boldsymbol{\boldsymbol{\nabla}}\mathbf{v-}\boldsymbol{\boldsymbol{\nabla}%
}\cdot\left(  \kappa\boldsymbol{\boldsymbol{\nabla}}T\right)  =0.
\]
The first term in this equation describes production of heat due to
compressibility and viscosity, and the second term describes redistribution
(diffusion) of the produced heat.

The asymptotic form of the \underline{velocity} equation depends on whether or
not $\mathbf{r}$ is close to a wall.

First, consider the \underline{outer} region, i.e., far from walls, where the
wall-induced potential $U(\mathbf{r})$ can be neglected and the fluid--fluid
interaction term can be rearranged as follows:%
\begin{multline*}
\int_{\mathcal{D}}\rho(\mathbf{r}_{1},t)\,\Phi(\varepsilon^{-1}\left\vert
\mathbf{r-r}_{1}\right\vert )\,\mathrm{d}^{3}\mathbf{r}_{1}\\
=\varepsilon^{3}\rho(\mathbf{r},t)\int\Phi(r_{1})\,\mathrm{d}^{3}%
\mathbf{r}_{1}\\
+\varepsilon^{5}\left[  \nabla^{2}\rho(\mathbf{r},t)\right]  \int r_{1}%
^{2}\,\Phi(r_{1})\,\mathrm{d}^{3}\mathbf{r}_{1}+\mathcal{O}(\varepsilon^{7}).
\end{multline*}
Taking into account (\ref{2.24}), and omitting the small terms, one can
rewrite Eq. (\ref{2.21}) in the form%
\begin{equation}
\frac{1}{\rho}\,\boldsymbol{\boldsymbol{\nabla}}\cdot\left(  \mathbf{I}%
p-\boldsymbol{\Pi}\right)  =\boldsymbol{\boldsymbol{\nabla}}\nabla^{2}\rho,
\label{3.2}%
\end{equation}
or equivalently%
\begin{equation}
\boldsymbol{\boldsymbol{\nabla}}\cdot\left(  \mathbf{I}p-\boldsymbol{\Pi
}\right)  =\boldsymbol{\boldsymbol{\nabla}}\cdot\left[  \mathbf{I}\left(
\rho\nabla^{2}\rho+\frac{1}{2}\left\vert \mathbf{\nabla}\rho\right\vert
^{2}\right)  -\left(  \boldsymbol{\boldsymbol{\nabla}}\rho\right)  \left(
\boldsymbol{\boldsymbol{\nabla}}\rho\right)  \right]  ,\nonumber
\end{equation}
where the expression in the square brackets is the Korteweg stress.

Most importantly, the fluid--fluid interaction term in Eq. (\ref{3.2}) is
differential -- hence, a boundary condition for $\rho$ is needed. It will be
derived by matching the outer solution to that in the inner (near-wall) region.

Next, consider the \underline{inner} region of characteristic thickness
$\varepsilon$. Assuming for simplicity that the wall passes through the origin
of the coordinate system and is tangent to the $\left(  x,y\right)  $ plane
(so that $U$ depends locally only on $z$), one can neglect the curvature of
the inner layer and introduce the inner coordinate $\hat{z}=\varepsilon^{-1}%
z$. The dependence on $x$ and $y$, in turn, is forced by the outer flow --
hence, these variables do not need rescaling except when they appear in $\Phi
$, which depends on $\hat{r}=\left(  \hat{x}^{2}+\hat{y}^{2}+\hat{z}%
^{2}\right)  ^{1/2}$ where $\hat{x}=\varepsilon^{-1}x$ and $\hat
{y}=\varepsilon^{-1}y$.

As for the unknowns, the density does not need rescaling, but the velocity
does, as the no-flow boundary condition (\ref{2.18}) suggest $\mathbf{\hat{v}%
}=\mathbf{v}/\varepsilon$. Finally, since the inner region is thin, the
temperature there should be assumed to be independent of $\hat{z}$.

One can see that the rescaled version of Eq. (\ref{2.21}) is dominated by the
fluid--fluid and solid--fluid interactions. Thus, to leading order, one
obtains%
\begin{equation}
\frac{\partial}{\partial\hat{z}}\left[  \int_{0}^{\infty}\rho(\hat{z}%
_{1})\,\Psi(\hat{z}-\hat{z}_{1})\,\mathrm{d}\hat{z}_{1}+U(\hat{z})\right]  =0,
\label{3.3}%
\end{equation}
where%
\begin{equation}
\Psi(\hat{z})=\int\int\Phi(\hat{r})\,\mathrm{d}\hat{x}\,\mathrm{d}\hat{y}.
\label{3.4}%
\end{equation}
Observe that condition (\ref{2.23}) and the symmetry of $\Phi(\hat{r})$ imply%
\begin{equation}
\int\Psi(\hat{z})\,\mathrm{d}\hat{z}=0,\qquad\int\hat{z}\,\Psi(\hat
{z})\,\mathrm{d}\hat{z}=0. \label{3.6}%
\end{equation}
Given (\ref{3.6}), one can readily verify that Eq. (\ref{3.3}) is consistent
with the following long-range behavior:%
\begin{equation}
\hat{\rho}(\hat{z})\sim\rho_{0}+\rho_{0}^{\prime}\,\hat{z}+\frac{1}{2}\rho
_{0}^{\prime\prime}\,\hat{z}^{2}\qquad\text{as}\qquad\hat{z}\rightarrow\infty,
\label{3.7}%
\end{equation}
where $\rho_{0}$, $\rho_{0}^{\prime}$, and $\rho_{0}^{\prime\prime}$ do not
depend on $\hat{z}$ (but can depend on $x$, $y$, and $t$).

Asymptotic (\ref{3.7}) is to be matched to the outer solution. If $\rho
_{0}^{\prime\prime}\neq0$, (\ref{3.7}) implies that the outer solution is such
that%
\[
\frac{\partial^{2}\rho}{\partial z^{2}}=O(\varepsilon^{-2})\qquad
\text{as}\qquad z\rightarrow0,
\]
indicating a mismatch unless $\rho_{0}^{\prime\prime}=0$. A similar argument
yields $\rho_{0}^{\prime}=0$, so that the boundary condition for the outer
solution is%
\begin{equation}
\rho=\rho_{0}\qquad\text{at}\qquad\mathbf{r}\in\partial\mathcal{D}.
\label{3.8}%
\end{equation}
The parameter $\rho_{0}$ should be calculated by solving Eq. (\ref{3.3})
subject to condition (\ref{3.7}) with $\rho_{0}^{\prime}=\rho_{0}%
^{\prime\prime}=0$. Physically, $\rho_{0}$ is determined by a balance between
the solid--fluid and fluid--fluid interactions.

An example where Eq. (\ref{3.3}) can be solved analytically is given in
Appendix \ref{Appendix B}.

\subsection{Static interfaces}

The rest of this work is concerned with static interfaces, for which
$\partial/\partial t=0$ and $\mathbf{v}=\mathbf{0}$. It is also clear that
static fluid ought to be isothermal (in all models, both exact and
asymptotic), so $T=\operatorname{const}$.

Taking this into account and returning to the dimensional variables, one can
write the asymptotic equations (\ref{3.2}) in the form%
\begin{equation}
\frac{1}{\rho}\,\boldsymbol{\boldsymbol{\nabla}}%
p=K\boldsymbol{\boldsymbol{\nabla}}\nabla^{2}\rho, \label{3.9}%
\end{equation}
where it is implied that $p$ depends on $\rho$ and, parametrically, on $T$.
Equation (\ref{3.9}) and the boundary condition (\ref{3.8}) fully determine
$\rho(\mathbf{r})$.

Eq. (\ref{3.9}) can be rewritten in a mathematically equivalent (but, in some
cases, more convenient) form. Multiplying (\ref{3.9}) by $\rho$ and
integrating, one obtains%
\begin{equation}
p-p_{0}=K\left(  \rho\nabla^{2}\rho-\frac{1}{2}\left\vert
\boldsymbol{\boldsymbol{\nabla}}\rho\right\vert ^{2}\right)  , \label{3.10}%
\end{equation}
where $p_{0}$ is a constant of integration (and, physically, the pressure at
infinity). Alternatively, using identity (\ref{2.6}), one can rewrite
(\ref{3.9}) in terms of the Gibbs free energy,%
\begin{equation}
G-G_{0}=K\nabla^{2}\rho, \label{3.11}%
\end{equation}
where $G_{0}$ is a constant of integration (and, physically, the equilibrium
value of $G$).

\section{The surface tension vs. temperature\label{Sect. 4}}

\subsection{Theory}

It is well known (e.g., \cite{Mauri13}) that the surface tension of a
liquid--vapor interface can be related to the one-dimensional solution of Eq.
(\ref{3.9}). To do so, substitute $\rho=\rho_{lv}(z)$ into (\ref{3.9}) which
yields%
\begin{equation}
\frac{1}{\rho_{lv}}\frac{\mathrm{d}p(\rho_{lv},T)}{\mathrm{d}z}=K\frac
{\mathrm{d}^{3}\rho_{lv}}{\mathrm{d}z^{3}}. \label{4.1}%
\end{equation}
There are no solid boundaries in this case, thus,%
\begin{align}
\rho &  \rightarrow\rho_{l}\qquad\text{as}\qquad z\rightarrow-\infty
,\label{4.2}\\
\rho &  \rightarrow\rho_{v}\qquad\text{as}\qquad z\rightarrow+\infty,
\label{4.3}%
\end{align}
where $\rho_{l}$ and $\rho_{v}$ are the densities of the liquid and vapor,
respectively. Using the one-dimensional reductions of Eqs. (\ref{3.10}%
)-(\ref{3.11}), one can show that the boundary-value problem (\ref{4.1}%
)-(\ref{4.3}) has a solution only if $\rho_{l}$ and $\rho_{v}$ satisfy the
Maxwell construction, i.e., the following algebraic equations:%
\begin{equation}
p(\rho_{l},T)=p(\rho_{v},T),\qquad G(\rho_{l},T)=G(\rho_{v},T). \label{4.4}%
\end{equation}
Eqs. (\ref{4.4}) determine how $\rho_{l}$ and $\rho_{v}$ depend on $T$;
interestingly, they are exact despite the approximate nature of the DIM.

Once the boundary-value problem (\ref{4.1})-(\ref{4.4}) is solved and its
solution $\rho_{lv}(z)$ is found, the surface tension is given by%
\begin{equation}
\sigma=K\int_{-\infty}^{\infty}\left(  \frac{\mathrm{d}\rho_{lv}}{\mathrm{d}%
z}\right)  ^{2}\mathrm{d}z. \label{4.5}%
\end{equation}

\subsection{Comparison with observations}

Before comparing the dependence of $\sigma$ on $T$ determined by (\ref{4.5})
to that measured for a specific fluid, one has to specify the EoS and the
Korteweg parameter $K$. The former will be approximated by the Enskog--Vlasov
EoS (see Appendix \ref{Appendix A}) and the latter is discussed below.

\begin{figure}
\includegraphics[width=\columnwidth]{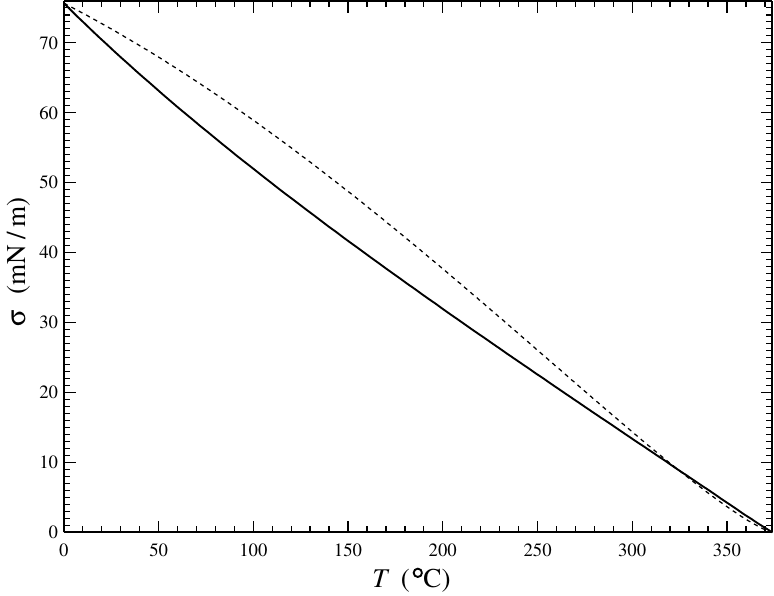}
\caption{The surface tension $\sigma$ of the interface between liquid water and its vapor vs. the temperature $T$. The temperature varies from water's triple-point value to its critical value. The solid curve shows the results computed through the DIM, the dotted curve shows the corresponding empiric results \cite{WagnerKretzschmar08}.}
\label{fig1}
\end{figure}

The simplest way to fix $K$ consists in solving the boundary-value problem
(\ref{4.1})-(\ref{4.3}) for a certain value of $T$ -- say, at the triple point
-- and ensure that the value of $\sigma$ predicted by (\ref{4.5}) coincides
with the surface tension $\sigma_{r}$ measured for a real liquid--vapor
interface\footnote{If, for a fluid under consideration, the surface tension of
the liquid--vapor interface has never been measured, but has been computed
through (presumably highly accurate) molecular and Monte Carlo simulations,
one can benchmark the prediction of Eq. (\ref{4.5}) against the results of the
latter (as done in Ref. \cite{GalloMagalettiCasciola18}).}. For water, the
latter value is \cite{WagnerKretzschmar08}%
\begin{equation}
\sigma_{r}=75.65\times10^{-3}\mathrm{N~m}^{-1}\hspace{0.5cm}\text{at}%
\hspace{0.5cm}T=273.16~\mathrm{K}.\label{4.6}%
\end{equation}
Eqs. (\ref{4.1})-(\ref{4.4}) with the EoS given by (\ref{2.8}), (\ref{A.1}%
)-(\ref{A.5}) were solved numerically and the computed $\rho_{lv}(z)$ was
substituted into (\ref{4.5}). The resulting $\sigma$ agrees with (\ref{4.6})
if%
\begin{equation}
K=2.45\times10^{-17}\mathrm{m}^{7}\mathrm{kg}^{-1}\mathrm{N}^{-2}.\label{4.7}%
\end{equation}
Now, one can compute $\sigma(T)$ for the whole temperature range where liquid
water and vapor coexist, i.e., between the triple and critical points. The
theoretical dependence is compared to the empiric one in Fig. \ref{fig1}:
evidently, the two sets of results agree well.

\subsection{Discussion: the width of a liquid--vapor
interface\label{Sect. 4.3}}

It is instructive to consider the boundary-value problem (\ref{4.1}%
)-(\ref{4.4}) in the small-temperature limit. Assuming the Enskog--Vlasov EoS
(\ref{2.8}) and omitting the term involving $T$, one can write (\ref{4.1}) in
the form%
\begin{equation}
-2a\frac{\mathrm{d}\rho_{lv}}{\mathrm{d}z}=K\frac{\mathrm{d}^{2}\rho_{lv}%
}{\mathrm{d}z^{2}}.\label{4.8}%
\end{equation}
At low $T$, the vapor density is negligible, whereas the liquid density is
close to its triple-point density -- so that the boundary conditions
(\ref{4.2})-(\ref{4.3}) become%
\begin{align}
\rho &  \rightarrow\rho_{tp}\qquad\text{as}\qquad z\rightarrow-\infty
,\label{4.9}\\
\rho &  \rightarrow0\hspace{0.98cm}\text{as}\qquad z\rightarrow+\infty
,\label{4.10}%
\end{align}
The solution of the boundary-value problem (\ref{4.8})-(\ref{4.10}) is%
\[
\rho=\left\{
\begin{tabular}
[c]{ll}%
$\rho_{tp}$\medskip & if$\qquad z\leq-\frac{1}{2}W,$\\
$\tfrac{1}{2}\rho_{tp}\left(  1-\sin\dfrac{\pi z}{W}\right)  \qquad$\medskip &
if$\qquad\left\vert z\right\vert <~\frac{1}{2}W,$\\
$0$ & if$\qquad z\geq~~~\frac{1}{2}W,$%
\end{tabular}
\ \right.
\]
where%
\begin{equation}
W=\pi\left(  \frac{K}{2a}\right)  ^{1/2}\label{4.11}%
\end{equation}
is, physically, the low-$T$ limit of the width of the interface. Expression
(\ref{4.11}) agrees qualitatively with the estimate of the interfacial
thickness obtained in Refs.
\cite{MagalettiGalloMarinoCasciola16,GalloMagalettiCoccoCasciola20}: if
adapted for the Enskog--Vlasov EoS and $T=0$, the latter yields a result which
is $\pi$ times smaller than (\ref{4.11}).

Substituting estimates (\ref{4.7}) for $K$ and (\ref{A.1}) for $a$ into
expression (\ref{4.11}), one obtains%
\begin{equation}
W\approx2.40\times10^{-10}\mathrm{m}.\label{4.12}%
\end{equation}
It is also instructive to estimate the characteristic intermolecular distance
$D$ for liquid water -- at, say, the triple point:
\[
D\approx n_{tp}^{-1/3}\approx3.11\times10^{-10}\mathrm{m},
\]
where $n_{tp}$ is the triple-point number density.

Thus, for small $T$, the thickness of the liquid--vapor interface is
comparable to the intermolecular distance. With increasing $T$, estimate
(\ref{4.12}) becomes invalid, as thermal motion of molecules erodes the
interface, making it thicker. Finally, when $T$ approaches the critical point,
the liquid--vapor interface becomes much thicker than $D$.

Note that the DIM is not the first hydrodynamic model to be used at scales
comparable to $D$, where it is not formally applicable. The standard
Navier-slip boundary condition -- routinely used in almost all studies of
contact lines -- implies the same. The justification of using hydrodynamic
models at small scales is as follows: even though they cannot accurately
predict the microscopic characteristics of interfaces, the structure of those
is still qualitatively correct -- as is (sic!) their effect on the macroscopic
flow. This appears to be true for the DIM, which predicts the correct
macroscopic properties of fluids in equilibrium (the Maxwell construction), as
well as their surface tension.

Note also that small interfacial thickness might hamper applications of the DIM with a
realistic EoS to numerical modeling of contact lines. One should still be able
to use it in conjunction with the numerical techniques recently developed for
nucleation and collapse of vapor bubbles
\cite{MagalettiMarinoCasciola15,MagalettiGalloMarinoCasciola16,GalloMagalettiCasciola18}
and drops impacting on a solid surface
\cite{GelissenVandergeldBaltussenKuerten20}.

\section{The contact angle vs. temperature\label{Sect. 5}}

To define the contact angle, one needs to introduce the boundary-value
problems describing solid--liquid and solid--vapor interfaces (the same way
problem (\ref{4.1})-(\ref{4.3}) describes liquid--vapor interfaces). To do so,
introduce $\rho_{sl}(z)$ and $\rho_{sv}(z)$ satisfying%
\begin{equation}
\frac{1}{\rho_{sl}}\frac{\mathrm{d}p(\rho_{sl},T)}{\mathrm{d}z}=K\frac
{\mathrm{d}^{3}\rho_{sl}}{\mathrm{d}z^{3}}, \label{5.1}%
\end{equation}%
\begin{align}
\rho_{sl}  &  =\rho_{0}\hspace{0.73cm}\text{at}\qquad z=0,\label{5.2}\\
\rho_{sl}  &  \rightarrow\rho_{l}\qquad\text{as}\qquad z\rightarrow+\infty,
\label{5.3}%
\end{align}
and%
\begin{equation}
\frac{1}{\rho_{sv}}\frac{\mathrm{d}p(\rho_{sv},T)}{\mathrm{d}z}=K\frac
{\mathrm{d}^{3}\rho_{sv}}{\mathrm{d}z^{3}}, \label{5.4}%
\end{equation}%
\begin{align}
\rho_{sv}  &  =\rho_{0}\hspace{0.8cm}\text{at}\qquad z=0,\label{5.5}\\
\rho_{sv}  &  \rightarrow\rho_{v}\qquad\text{as}\qquad z\rightarrow+\infty,
\label{5.6}%
\end{align}
where $\rho_{l}$ and $\rho_{v}$ are determined by the Maxwell construction
(\ref{4.4}).

Next, let the solid surface coincide with the $\left(  x,y\right)  $ plane and
the contact line, with the $y$ axis. This setting is described by the
two-dimensional version of (\ref{3.11}),%
\[
G-G_{0}=K\left(  \frac{\partial^{2}\rho}{\partial x^{2}}+\frac{\partial
^{2}\rho}{\partial z^{2}}\right)  ,
\]
and the following boundary conditions:%
\[
\rho=0\qquad\text{at}\qquad z=0,
\]%
\begin{equation}
\rho\rightarrow\rho_{sg}(z)\qquad\text{as}\qquad x\rightarrow-\infty,
\label{5.7}%
\end{equation}%
\begin{equation}
\rho\rightarrow\rho_{sl}(z)+\rho_{lv}(z\cos\theta-x\sin\theta)\qquad
\text{as}\qquad x\rightarrow+\infty, \label{5.8}%
\end{equation}
where the contact angle $\theta$ is implied to be less than $90^{\circ}$
(i.e., the solid is hydrophilic). As shown in Ref. \cite{PismenPomeau00}, one
can find $\theta$ without solving the above boundary-value problem:%
\begin{multline}
\left[
{\displaystyle\int_{-\infty}^{\infty}}
\left(  \dfrac{\mathrm{d}\rho_{lv}}{\mathrm{d}z}\right)  ^{2}\mathrm{d}%
z\right]  \cos\theta\\
=%
{\displaystyle\int_{0}^{\infty}}
\left(  \dfrac{\mathrm{d}\rho_{sv}}{\mathrm{d}z}\right)  ^{2}\mathrm{d}z-%
{\displaystyle\int_{0}^{\infty}}
\left(  \dfrac{\mathrm{d}\rho_{sl}}{\mathrm{d}z}\right)  ^{2}\mathrm{d}z.
\label{5.9}%
\end{multline}
If $\theta>90^{\circ}$ (hydrophobic solids), the boundary conditions
(\ref{5.7})-(\ref{5.8}) need to be slightly modified, but expression
(\ref{5.9}) remains exactly the same.

\begin{figure}
\includegraphics[width=\columnwidth]{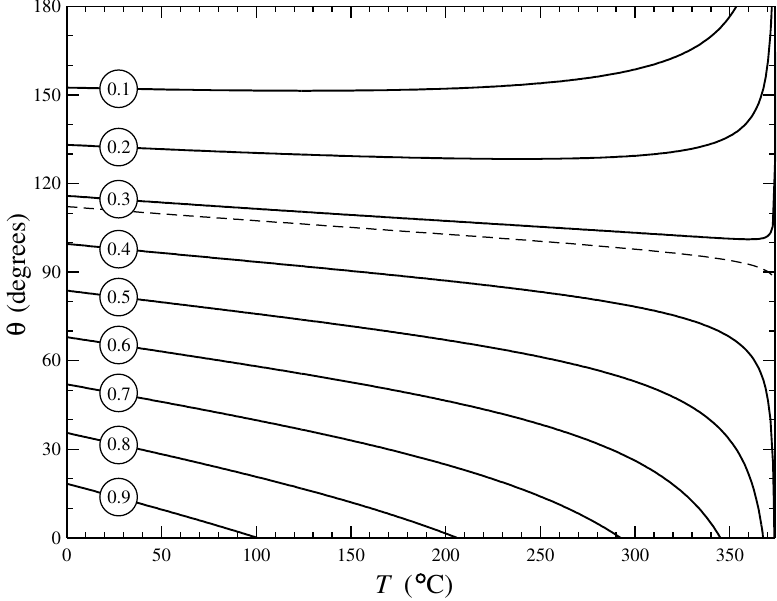}
\caption{The static contact angle $\theta$ vs. the temperature $T$, for water. The curves are labelled with the corresponding values of $\rho _{0}/\rho_{tp}$ where $\rho_{tp}$ is the triple-point density of liquid water. The separatrix (dashed line) corresponds to $\rho_{0}$ coinciding with the critical density.}
\label{fig2}
\end{figure}

Thus, $\theta$ can be computed by integrating the boundary-value problems
(\ref{5.1})-(\ref{5.3}) and (\ref{5.4})-(\ref{5.6}) numerically and
substituting their solutions into expression (\ref{5.9}). The solution of
problem (\ref{4.1})-(\ref{4.3}) is not needed, as it can be readily shown that%
\[%
{\displaystyle\int_{-\infty}^{\infty}}
\left(  \dfrac{\mathrm{d}\rho_{lv}}{\mathrm{d}z}\right)  ^{2}\mathrm{d}z=%
{\displaystyle\int_{0}^{\infty}}
\left(  \dfrac{\mathrm{d}\rho_{sv}}{\mathrm{d}z}\right)  ^{2}\mathrm{d}z+%
{\displaystyle\int_{0}^{\infty}}
\left(  \dfrac{\mathrm{d}\rho_{sl}}{\mathrm{d}z}\right)  ^{2}\mathrm{d}z.
\]
Unfortunately, there seems to be no measurements of the contact angle of a
single-fluid interface, on a substrate with a sufficiently narrow hysteresis
interval. Thus, instead of examining a specific water--substrate combination,
$\theta$ was computed for the full range of $\rho_{0}$ -- from zero to the
triple-point density. The results are presented in Fig. \ref{fig2}.

Evidently, for all $\rho_{0}$ except a certain separatrix value, a temperature
exists such that the contact angle either becomes$\ $equal to $180^{\circ}$
(perfect hydrophobicity) or $0^{\circ}$ (perfect hydrophilicity). The former
occurs if $\rho_{0}$ matches the saturated-vapor density $\rho_{v}(T)$ and the
latter, if $\rho_{0}$ matches the liquid density $\rho_{l}(T)$. It is also
clear that the separatrix corresponds to $\rho_{0}$ equal to the critical density.

\section{Summary and concluding remarks}

Thus, the following results have been obtained:

\begin{enumerate}
\item It has been shown that the boundary condition prescribing the density of
a fluid at a solid wall can be derived without assuming that the solid--fluid
interaction is short-ranged by comparison with the fluid--fluid one (as
conjectured in Ref. \cite{PismenPomeau00}). Thus, this boundary condition is
based on the same physics as the DIM itself.

\item A parameter region has been identified [$\beta\ll1$ with $\beta$
determined by (\ref{2.23})], where interfacial flows without external heating
are almost isothermal. This region does \emph{not} include water, where the
heat production due to viscosity and compressibility of vapor near the
interface is too strong.\\*\hspace*{0.6cm}It is worth mentioning that
interfaces are likely to be isothermal in fluids with high viscosity, such as
glycerol or ethylene glycol. This claim is supported by the estimates carried
out in Ref. \cite{Benilov20} using the van der Waals EoS; even though it is
much less accurate than the Enskog--Vlasov EoS used in the present work, it
still works qualitatively correct for water. Physically, high viscosity slows
the flow down and, thus, reduces the heat production.

\item The DIM was coupled with a realistic EoS of water and used to compute
the surface tension $\sigma$ of a liquid/vapor interface as a function of the
temperature $T$ (Sect. \ref{Sect. 4}, Fig. \ref{fig1}). The theoretical
results agree well with the empiric dependence $\sigma(T)$.

\item The static contact angle $\theta$ of a liquid--vapor interface has been
computed as a function of $T$ for water (Sect. \ref{Sect. 5}, Fig.
\ref{fig2}). The results obtained predict that, with increasing $T$, any
substrate would become either perfectly hydrophobic ($\theta=180^{\circ}$) or
perfectly hydrophilic ($\theta=0^{\circ}$).
\end{enumerate}

Admittedly, (the most counter-intuitive) conclusion 4 has not been verified experimentally.

To do so in the future, one needs to experiment with a \emph{single}-fluid
interface and a \emph{chemically-cleaned} or \emph{lubricant-impregnated}
substrate. The former requirement can be relaxed if the present results are
extended to a mixture of fluids (e.g., water plus nitrogen). The latter
requirement is crucial, however, as, for usual substrates, $\theta$ does not
assume a reasonably well-defined value, but one from an often-wide hysteresis interval.

Still, there is qualitative evidence that the effects of induced
hydrophobicity and hydrophilicity do occur in the real world.

It can be argued that states with $\theta=180^{\circ}$ or $\theta=0^{\circ}$
can be created through \emph{any} parameter variation, not only that of the
temperature. To do so, this variation should change either $\rho_{0}$ or the
densities of the phases -- until the former coincides with one of the latter:
perfect hydrophobicity and hydrophilicity correspond to $\rho_{0}=\rho_{v}$
and $\rho_{0}=\rho_{l}$, respectively. This argument could explain the
observed behavior of droplets under variable electric field
\cite{BrabcovaMcHaleWellsBrown17}.

Finally, note that a case has been made \cite{Pismen01,YochelisPismen06} for
switching from the simplified differential representation for the van der
Waals force (used in this paper and all applications) to the full integral
expression (\ref{2.9}). It is not clear at this stage how this would effect
our results.

\appendix

\section{The Enskog--Vlasov equation of state\label{Appendix A}}

When applying the DIM to a specific fluid, one needs an EoS describing this
fluid's thermodynamic properties with a reasonable accuracy. In this work, the
\emph{Enskog--Vlasov} (EV) model will be used, where the internal energy and
entropy per unit mass are given by Eqs. (\ref{2.7}), and the EoS, by
(\ref{2.8}). Note that Eqs. (\ref{2.7})-(\ref{2.8}) are invariant with respect
to a simultaneous change%
\[
b\rightarrow\operatorname{const}\times b,\qquad\Theta(\xi)\rightarrow
\Theta(\operatorname{const}^{-1}\times\xi).
\]
Thus, to remove the ambiguity when choosing $b$, the restriction%
\[
\left[  \frac{\mathrm{d}\Theta(\xi)}{\mathrm{d}\xi}\right]  _{\xi=0}%
=\frac{2\pi}{3}%
\]
is traditionally imposed in the EV theory.

\begin{figure}
\includegraphics[width=\columnwidth]{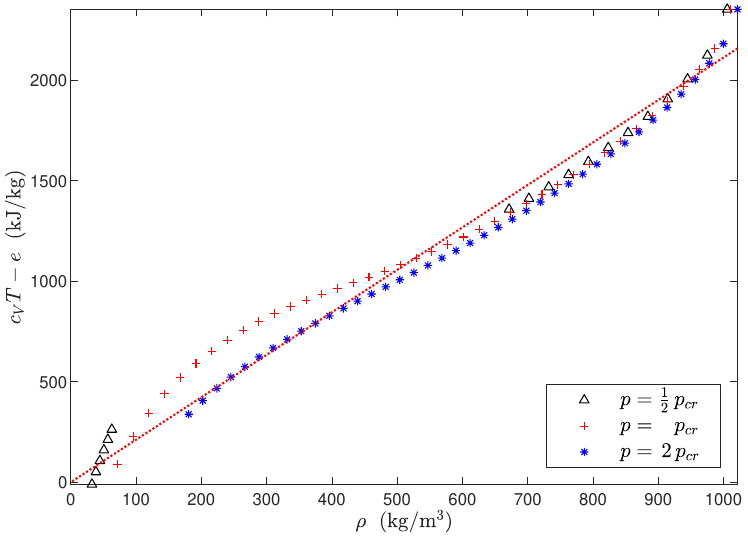}
\caption{The non-ideal component of the internal energy vs. density. The non-connected symbols show the empiric data from Ref. \cite{LinstromMallard97} presented in isobaric form, for the values of the pressure $p$ (relative to the critical pressure $p_{cr}$) stated in the
legend. The dotted line shows the linear fit of the critical isobar.}
\label{fig3}
\end{figure}

Before using EoS (\ref{2.8}), one needs to calibrate it, i.e., specify $a$,
$b$, and $\Theta(\xi)$, such that the fluid under consideration is described
as accurately as possible. As for the specific heat capacity, it will be
assigned the ideal-fluid value: for water, this amounts to%
\[
c_{V}=3R.
\]
To fix $a$, observe that, as follows from (\ref{2.7}),%
\[
\Delta e=c_{V}T-e
\]
depends linearly on $\rho$. Thus, $a$ can be determined by fitting a linear
function to the empiric dependence $\Delta e$ vs. $\rho$. Using the data from
Ref. \cite{LinstromMallard97}, one can estimate%
\begin{equation}
a=2112~\mathrm{m}^{5}\mathrm{s}^{-2}\mathrm{kg}^{-1}. \label{A.1}%
\end{equation}
(For simplicity, this estimate was obtained using only the data for the
critical pressure $p=220.64~\mathrm{bar}$ and the temperature range
$273.16~\mathrm{K}$ to $800.16~\mathrm{K}$.) The accuracy of representation
(\ref{2.7}) of the free energy can be assessed from Fig. \ref{fig3} which
shows the dependence $\Delta e$ vs. $\rho$ for three different isobars
(including the critical one), together with the linear fit resulting estimate
(\ref{A.1}).

\begin{figure}
\includegraphics[width=\columnwidth]{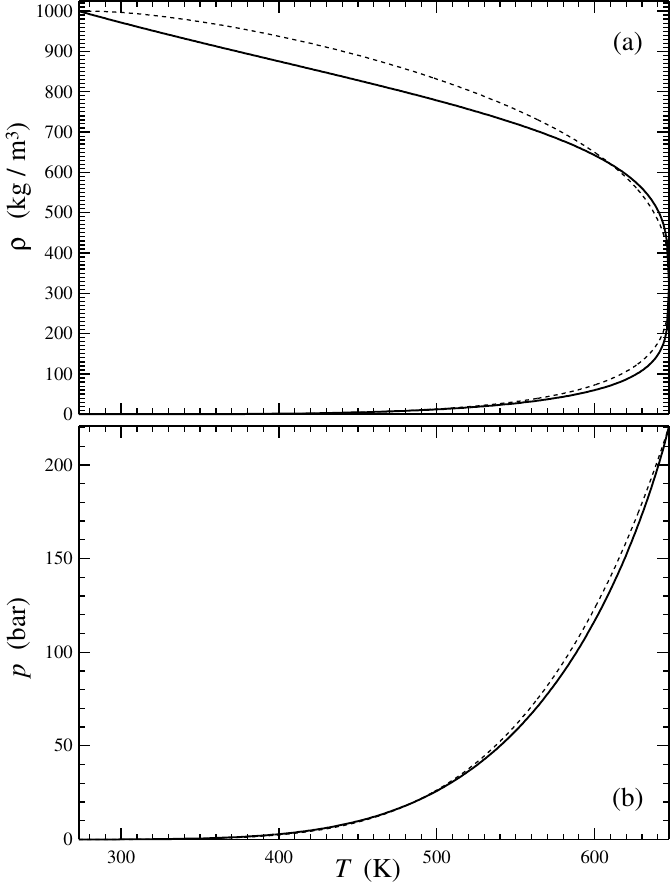}
\caption{The parameters of liquid water and vapor in equilibrium: the empiric data from Ref. \cite{LinstromMallard97} (dotted line) and the results obtained through the EV model (solid line). (a) The densities of the saturated vapor and liquid (the upper and lower parts of the curves, respectively) vs. $T$. (b) The pressure of the saturated vapor vs. $T$.}
\label{fig4}
\end{figure}

\begin{figure}
\includegraphics[width=\columnwidth]{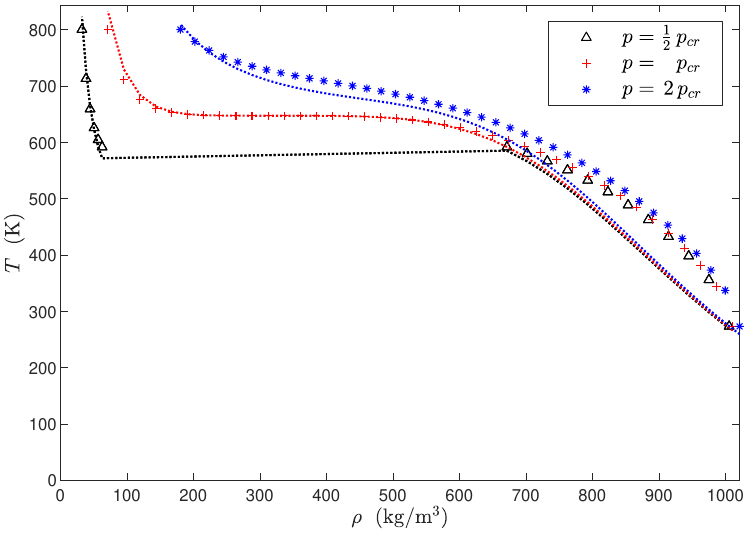}
\caption{The equation of state for water: the empiric data \cite{LinstromMallard97} (non-connected symbols) and the results obtained through the EV model (dotted curves).}
\label{fig5}
\end{figure}

The parameter $b$, in turn, can be simply equated to the reciprocal of the
triple-point density \cite{BenilovBenilov18,BenilovBenilov19} -- hence, for
water,%
\begin{equation}
b=1.0002\times10^{-3}\mathrm{m}^{3}\mathrm{kg}^{-1}. \label{A.2}%
\end{equation}
Finally, let%
\begin{equation}
\Theta(\xi)=\frac{2\pi}{3}\xi+\sum_{i=2}^{5}c_{i}\xi^{i}, \label{A.3}%
\end{equation}
with the coefficients $c_{i}$ being such that the equation of state
(\ref{2.8}), (\ref{A.1})-(\ref{A.3}) describes correctly the fluid's density
and temperature at the triple and critical points, as well as the critical
pressure (for more details, see Ref. \cite{BenilovBenilov19}). In application
to water, this yields%
\begin{align}
c_{2}  &  =\hspace{0.45cm}4.649,\qquad c_{3}=1.642,\label{A.4}\\
c_{4}  &  =-10.108,\qquad c_{5}=7.973. \label{A.5}%
\end{align}
The accuracy of the EV model calibrated this way can be assessed from Figs.
\ref{fig4}-\ref{fig5}, which compare predictions of (\ref{2.8}),
(\ref{A.1})-(\ref{A.5}) to the corresponding empiric results
\cite{LinstromMallard97}.

One can see, that the Enskog--Vlasov EoS is reasonably accurate and can be
safely used in studies of flows with phase transitions.

\section{An example of solution of Eq. (\ref{3.3})\label{Appendix B}}

The solution of the inner-problem equation (\ref{3.3}) will be illustrated by
the simplest particular case of $\Psi$ and $U$, such that the former is
approximated by a piece-wise-constant function%
\begin{equation}
\Psi(\hat{z})=\left\{
\begin{tabular}
[c]{ll}%
$0\smallskip$ & if$\hspace{0.3cm}\hat{z}\in\left(  -\infty,-2H\right)  ,$\\
$-1\hspace{0.2cm}\smallskip$ & if$\hspace{0.3cm}\hat{z}\in\left[
-2H,-H\right)  ,$\\
$2\smallskip$ & if$\hspace{0.3cm}\hat{z}\in\left[  -H,H\right]  ,$\\
$-1\smallskip$ & if$\hspace{0.3cm}\hat{z}\in\left(  H,2H\right]  ,$\\
$0$ & if$\hspace{0.3cm}\hat{z}\in\left(  2H,\infty\right)  ,$%
\end{tabular}
\ \right.  \label{B.1}%
\end{equation}
and the latter is approximated by a piece-wise-linear function,%
\begin{equation}
U(\hat{z})=\left\{
\begin{tabular}
[c]{ll}%
$U_{0}\hat{z}-\left(  U_{0}+U_{1}\right)  \Delta\hspace{0.2cm}\smallskip$ &
if$\hspace{0.3cm}\hat{z}\in\left[  0,H\right]  ,$\\
$U_{1}\left(  \hat{z}-2\Delta\right)  \smallskip$ & if$\hspace{0.3cm}\hat
{z}\in\left(  H,2H\right]  ,$\\
$0$ & if$\hspace{0.3cm}\hat{z}\in\left(  2H,\infty\right)  ,$%
\end{tabular}
\ \right.  \label{B.2}%
\end{equation}
where $\Delta$, $U_{0}$, and $U_{1}$ are constants. As required, the above
$\Psi(\hat{z})$ satisfies restrictions (\ref{3.6}).

Substituting (\ref{B.1})-(\ref{B.2}) into Eq. (\ref{3.3}), omitting hats, and
introducing%
\[
\rho_{n}(z)=\rho(z+n\Delta)\qquad\text{if}\qquad z\in\left(  0,\Delta\right]
,
\]
one obtains%
\begin{align*}
2\rho_{1}-\rho_{2}+U_{0}  &  =0,\\
2\left(  \rho_{2}-\rho_{0}\right)  -\rho_{3}+U_{1}  &  =0,\\
2\left(  \rho_{n-1}-\rho_{n-3}\right)  -\rho_{n}+\rho_{n-4}  &  =0\qquad
\text{for}\qquad n\geq4.
\end{align*}
One can use these (recursive) equations to calculate several first terms --
then guess the general formula relating $\rho_{n}$ to $\rho_{0}$ and $\rho
_{1}$ -- then verify this formula by substitution -- and thus obtain%
\begin{multline}
\rho_{n}=\frac{n^{2}}{4}\left(  \rho_{1}-\rho_{0}+U_{0}+U_{1}\right) \\
+\frac{n}{2}\left(  \rho_{1}-U_{1}\right)  +\rho_{0}\qquad\text{for even
}n\geq0, \label{B.3}%
\end{multline}%
\begin{multline}
\rho_{n}=\frac{n^{2}}{4}\left(  \rho_{1}-\rho_{0}+U_{0}+U_{1}\right)
+\frac{n}{2}\left(  \rho_{1}-U_{1}\right) \\
+\frac{1}{4}\left(  \rho_{1}+\rho_{0}-U_{0}+U_{1}\right)  \qquad\text{for odd
}n\geq1. \label{B.4}%
\end{multline}
Observe that the quadratic dependence of $\rho$ on $n$ is in line with that of
$\rho$ on $z$ in asymptotic (\ref{3.7}).

As shown in the main body of the paper, the inner solution matches the outer
one only if the former does not grow as $z\rightarrow\infty$. Thus, the
growing terms in expressions (\ref{B.3})-(\ref{B.4}) should be eliminated,
which implies $\rho_{0}=U_{0}$, $\rho_{1}=-U_{1}$, and%
\[%
\begin{tabular}
[c]{ll}%
$\rho_{n}=U_{0}+2U_{1}\qquad$ & $\text{for even }n\geq0,$\\
$\rho_{n}=U_{1}$ & $\text{for odd }n\geq1.$%
\end{tabular}
\]
This solution is bounded, but it oscillates, so still does not have the
desired (uniform) asymptotics as $z\rightarrow\infty$. The only way to
eliminate the oscillations is to require that $U_{0}=-U_{1}$ -- in which case
$\rho_{n}=U$ for all $n$ -- hence, $\rho(z)=U_{1}$ for all $z$, and%
\[
\rho_{0}=U_{1}.
\]
The fact that the near-wall region can generate short-scale oscillations and
potentially `send' them (through the matching conditions) into the whole
domain is interesting from the mathematical viewpoint. Physically, however,
such cases should be avoided, just like one of them has been avoided in the
above example.

In general, one can show that the large-$z$ asymptotics of the solution of Eq.
(\ref{3.3}) has a periodic component only if the Fourier transform of
$\Psi(z)$,%
\[
\hat{\chi}(k)=\int_{0}^{\infty}\Psi(z)\cos kz\,\mathrm{d}z,
\]
vanishes at some $k$. One can also show that the periodic component disappears
if the Fourier transform of $U(z)$ vanishes at the same value(s) of $k$ (which
is what happens in the above example when the condition $U_{0}=-U_{1}$ was applied).

\bibliography{.././../bib/refs}

\end{document}